\documentclass[aps,prl,twocolumn,groupedaddress]{revtex4-2}

\usepackage{amsmath}
\usepackage{amssymb}
\usepackage{amsfonts}
\usepackage[dvipsnames]{xcolor}
\usepackage{hyperref}
\hypersetup{
	pdftex,
	pdfstartview=FitH,
	bookmarksnumbered=true,
	pagebackref,
	colorlinks,
	breaklinks,
	urlcolor=Maroon,
	linkcolor=Maroon,
	citecolor=Bittersweet
} 
\usepackage{graphicx}
\DeclareGraphicsExtensions{.pdf,.png,.jpg,.mps}
\usepackage{tikz}
\usetikzlibrary{matrix,decorations.pathmorphing,decorations.markings,arrows,positioning,intersections,mindmap,backgrounds}
\usetikzlibrary{positioning,calc} 

\usepackage{booktabs}
\usepackage{physics}
\usepackage{diagbox}

\usepackage{lmodern} 
\usepackage{scalerel,accents,trimclip}
\usepackage{CJK}

\usepackage{shuffle}
\usepackage{cleveref}

\newcommand{\zb}{\bar{z}}

\newcommand{\hs}{\hat{s}}
\newcommand{\htt}{\hat{t}}

\newcommand{\calD}{\hat{\mathcal{D}}}

\usepackage{graphicx}

\begin{document}


	\begin{CJK*}{UTF8}{gbsn}

	\title{Energy-Energy Correlator from the AdS Virasoro-Shapiro Amplitude}

    \author{Lecheng Ren (任乐成)$^{1}$}
    \email{lecheng.ren@qmul.ac.uk}
	\author{Bo Wang (王波)$^{2}$}
	\email{b\_w@zju.edu.cn}
    \author{Congkao Wen (温从烤)$^{1}$}
	\email{c.wen@qmul.ac.uk}
    
    \affiliation{{$^{1}$}Centre for Theoretical Physics, Department of Physics and Astronomy,\\
    Queen Mary University of London, London, E1 4NS, UK}
	\affiliation{\mbox{{$^2$}Zhejiang Institute of Modern Physics, School of Physics, Zhejiang University,} \\Hangzhou, Zhejiang 310058, China }
	\date{\today}

	\begin{abstract}
		We establish a precise formula relating the world-sheet integral of the AdS Virasoro-Shapiro amplitude to the energy-energy correlator (EEC) in $\mathcal{N}=4$ super Yang-Mills theory at strong coupling. This mapping allows us to evaluate the coefficients of the AdS curvature expansion of the EEC in terms of the world-sheet integral over a unit disk. To illustrate this idea, we explicitly compute the flat-space contribution and the first curvature correction to the EEC. Our results provide a rigorous description of the stringy energy flow, demonstrating how world-sheet correlators imprint themselves on collider observables and offering a potential template for effective string descriptions of energy correlators in general gauge theories.
	\end{abstract}

	\maketitle
	
	\end{CJK*}

\noindent{\bf Introduction.} Understanding the flow of energy in quantum field theory is a fundamental endeavor that bridges the gap between formal correlation functions and physical detectors.  In the context of Quantum Chromodynamics (QCD), Energy-Energy Correlators (EEC) have emerged as powerful observables for probing the dynamics of the strong interaction \cite{Basham:1978bw,Basham:1978zq} and the internal structure of jets \cite{Chen:2020vvp,Komiske:2022enw} (see \cite{Moult:2025nhu} for a recent comprehensive review). By weighting the cross-section with the energy of detected particles, the EEC offers an infrared-safe probe of the theory's asymptotic states. Formally, it is defined in terms of the correlation of light-ray operators \cite{Kravchuk:2018htv,Kologlu:2019mfz}. However, the complicated operator structure, running coupling effects and confinement  inherent to QCD make its non-perturbative structure difficult to analyze.

$\mathcal{N}=4$ super Yang-Mills theory (SYM) serves as the ideal laboratory for this endeavor. Despite being conformal, it captures the physical evolution of energy flow analogous to QCD. By tuning the 't Hooft coupling, 
one observes a transition from collimated ``jets" to a uniform distribution. At weak coupling, the EEC mirrors perturbative QCD \cite{Belitsky:2013ofa,Belitsky:2013xxa,Belitsky:2013bja,Henn:2019gkr} with power-law singularities from light-ray OPEs \cite{Kologlu:2019mfz,Chang:2020qpj,Chen:2020adz,Chen:2021gdk,Chen:2023zzh,Chen:2024nyc,Dixon:2019uzg,Korchemsky:2019nzm}. Conversely, at strong coupling, due to AdS/CFT \cite{Maldacena:1997re,Gubser:1998bc,Witten:1998qj} it is described by the supergravity approximation and becomes flat \cite{Hofman:2008ar}.

Despite our control over the weak and infinite-coupling limits, the intermediate regime remains elusive (see \cite{Dempsey:2025yiv} for recent work in this direction). This is the domain where the ``stringy" nature of the theory is manifest, which is dual to the finite size effects of strings. Understanding how these stringy features imprint themselves on the angular distribution of energy is a crucial step toward connecting the point-like descriptions of perturbative scattering with the extended objects of holographic models. Recent breakthroughs in bootstrapping the AdS Virasoro-Shapiro amplitude \cite{Alday:2023jdk,Alday:2023mvu} permit a precise analysis of the stringy dynamics. 

In this Letter, we present a rigorous derivation of the stringy energy flow from a world-sheet point of view. We establish a direct mapping between the world-sheet dynamics and the statistical energy correlations on the boundary. By transforming the AdS Virasoro-Shapiro amplitude from Mellin space to the detector sphere, we express the coefficients of the EEC in the AdS curvature expansion as integrals over the string moduli space. This result provides an explicit dictionary for translating world-sheet resonances into observable angular patterns. As a concrete application, we compute the flat-space limit and the first curvature correction, offering a prototype for constructing effective string descriptions of energy flow applicable to QCD-like theories.

\vspace{0.8em}

\noindent{\bf Energy-energy correlator from local correlator.} 
Consider a scattering process created by some source operator $\mathcal{O}$ with total momentum $q^{\mu}$, the EEC measures the energy flux correlation between two detectors, defined as
\begin{equation*}
    \text{EEC}(\xi) =  \frac{8 \pi^2}{q^2 \sigma_0} \int \mathrm{d}^4 x_{13} \, e^{iq\cdot  x_{13}} \langle \mathcal{O} (x_1)\mathcal{E}(n_2)\mathcal{E}(n_4)\mathcal{O}(x_3)\rangle \, ,
\end{equation*}
where $x_{ij}=x_{i}-x_j$ and $\sigma_0$ is the total cross-section. The energy flow operator is given by the integral of the stress-tensor along null infinity,
\begin{equation} 
\mathcal{E}(\vec{n}) = \lim_{r \to \infty} r^2 \int_{0}^{\infty} dt \, {n}_i\, T_{0i}(t, r\hat{n}) \,. 
\end{equation}
Here, $\mathcal{E}$ represents the energy flow operator placed at null infinity in the direction specified by the unit vector $n=(1,\hat{n})$ and $T_{\mu\nu}$ is the stress tensor. The correlator depends on a single cross ratio $\xi$ related to the angle $\theta$ between two detectors, in the rest frame of the source we have $\xi =  \sin^2(\theta/2)$. 

For $\mathcal{N}=4$ SYM, the EEC can be related to a four-point function of superconformal primary operators of dimension two 
(denoted $\mathcal{O}_{\bf 20'}$ in the literature). Crucially, these operators belong to the same supermultiplet as stress tensor, which makes it possible to relate their correlation function to the EEC. A key step is to Wick rotate the Euclidean correlation function -- where most explicit computations in the literature have been performed -- to the corresponding Wightman function. 
This procedure leads to the Mellin representation of the energy-energy correlator established in \cite{Belitsky:2013xxa,Belitsky:2013bja}   
\begin{equation}\label{eq:EECMellin}
    \text{EEC}(\xi)=  \int_{-i\infty}^{+i\infty} \frac{\mathrm{d}s }{2 \pi i }\frac{\mathrm{d}t}{2 \pi i }\, M(s,t) K(t,\xi) \, ,
\end{equation}
where the kernel is given by
\begin{equation}
    K(t,\xi) = -\frac{\pi \, t^2(t-2)^2}{128\,\xi^3\sin\left({\pi t}/{2}\right)} \left(\frac{\xi}{1-\xi}\right)^{1+\frac{t}{2}}  \, .
\end{equation}
We can choose the contour between $1<\text{Re}(s),\, \text{Re}(t)<2$. $M(s,t)$ denotes the Mellin amplitude of the four-point function $\mathcal{H}(u, v)$ of $\mathcal{O}_{\bf 20'}$'s 
\begin{align}
    \mathcal{H}(u, v) =& \int_{-i\infty}^{+i \infty} \frac{ \mathrm{d}s \, \mathrm{d}t }{(2\pi i)^2 } u^s v^t  M(s,t) \nonumber \\ 
    &\times \Gamma\left(2-s/2\right)^2 \Gamma(2-t/2)^2 \Gamma(2-\tilde{u}/2)^2\, ,
\end{align}
where $u, v$ are cross ratios and $s+t+\tilde{u}=4$. We will focus on the planar limit. In the strong coupling expansion, $M(s,t)$ comprises the supergravity contribution and an infinite series of stringy corrections (see e.g. \cite{Goncalves:2014ffa, Rastelli:2016nze, Binder:2019jwn, Drummond:2020dwr, Chester:2020dja, Abl:2020dbx, Aprile:2020mus}),
\begin{equation}
    M(s,t) = \frac{8}{(s-2)(t-2)(\tilde{u}-2)} +  \frac{120 \zeta_3}{\lambda^{3/2}} + \cdots \, .
\end{equation}
Higher-order corrections 
consist of symmetric polynomials in the Mellin variables $s$, $t$ and $\tilde{u}$. While substituting the leading supergravity term into \cref{eq:EECMellin} yields precisely 1 \cite{Hofman:2008ar}, a naive term-by-term integration of the higher-order polynomials leads to divergences in the $s$-integration. To obtain a well-defined result, we must reorganize the summation and integration. The physical EEC, which is expected to be free of divergence, can be organized in the strongly coupled regime as
\begin{align} \label{eq:EECxi}
    \text{EEC}(\xi)
    &=1+ \sum_{k=0}^{\infty} \lambda^{-\frac{k}{2}-1} \sum_{n=0}^{k} c_{k,n}  Q_{k-n}(\xi) \, . 
\end{align}
Here, the functions $Q_n(\xi)$ carry the dependence on the variable $t$ in the kernel, defined as \footnote{Note that we use a slightly different definition for $Q_n(\xi)$ compared to \cite{Goncalves:2014ffa}. In particular, $Q_n(\xi)$ defined here is equal to $Q_{n+1}(\xi)$ in \cite{Goncalves:2014ffa}.}
\begin{equation} \label{eq:Qn}
    Q_n(\xi)= \int_{-i\infty}^{+i\infty} \frac{\mathrm{d}t}{2 \pi i}\, t^{n} \, K(t,\xi) \, .
\end{equation}
It is easy to note $Q_{0}(\xi)=1 - 6 \xi + 6 \xi^2$, and more general $Q_n(\xi)$ can be determined through a recurrence relation $Q_{n}(\xi)={\partial_{\xi} (2 (1-\xi) \xi Q_{n-1}(\xi))}-2 (\xi-1) Q_{n-1}(\xi)$. 

The validity of the organization in \cref{eq:EECxi} is rooted in the structure of stringy corrections. The generic form of the stringy corrections to the Mellin amplitude scales as $M(s,t) \sim \lambda^{-\frac{3+k}{2}} \mathcal{P}_k(s,t)$, where $\mathcal{P}_k(s,t)$ is a polynomial of degree $k$. After integrating out $s$, the polynomials $\mathcal{P}_k(s,t)$ reduce to polynomials in $t$. This naturally organizes the expansion: terms with the highest power $t^k$ correspond to the index $n=0$ (i.e., $c_{k,0}$) and are governed strictly by the flat-space limit. Subleading powers $t^{k-1}$ correspond to $n=1$ (i.e., $c_{k,1}$) and receive contributions from the first curvature correction, and so forth. Thus, the index $n$ effectively counts the order in the curvature expansion away from the flat-space limit. Previous studies have determined that $c_{0,0}= 4\pi^2$ \cite{Hofman:2008ar, Goncalves:2014ffa}, as well as results for $c_{k,0}$ and specific coefficients such as $c_{1,1}$ obtained recently in \cite{Dempsey:2025yiv}. 
The objective of this Letter is to provide a universal algorithm for computing the general coefficients $c_{k,n}$. 

A rigorous derivation of these coefficients, which circumvents the ambiguities of the naive expansion, is achieved by establishing a connection between the EEC and the AdS Virasoro-Shapiro amplitude $A(S,T)$ \cite{Alday:2023jdk,Alday:2023mvu}. Since $A(S,T)$ is related to the Mellin amplitude via a Borel transform \cite{Penedones:2010ue}, we obtain
\begin{equation}\label{EECfromAdSVS}
    \text{EEC}(\xi)= \frac{1}{2\lambda^{3/2}} \hspace{-0.4em} \int_{0}^{\infty} \hspace{-0.4em}\mathrm{d \beta}  \int_{-i\infty}^{+i\infty} \hspace{-0.4em}\frac{\mathrm{d}s \, \mathrm{d}t}{(2 \pi i)^2}\, \frac{\beta^5}{e^{\beta}} A(\beta \hat{s}, \beta \hat{t}) K(t,\xi) \, ,
\end{equation}
where
\begin{equation} \label{eq:shth}
    \hat{s} = \frac{s-4/3}{2 \sqrt{\lambda}}\, , \qquad  \hat{t} = \frac{t-4/3}{2 \sqrt{\lambda}}\, .
\end{equation}
Recently, it was shown that the AdS Virasoro-Shapiro amplitude is expressed as a simple integral of a world-sheet correlator over the Riemann sphere \footnote{The similar construction is also valid for more holographic backgrounds such as \cite{Fardelli:2023fyq,Wang:2025pjo,Alday:2024yax,Alday:2024ksp,Wang:2025owf,Chester:2024wnb,Jiang:2025oar,Chester:2024esn}.},
\begin{equation}
    A(S,T) =\! \int \! \mathrm{d}^{2} z |z|^{-2S-2} |1{-}z|^{-2T-2} G^{tot}(z,\zb;S,T) \, ,
\end{equation}
where the measure reads $\mathrm{d}^2z =\frac{\mathrm{d} z\wedge \mathrm{d} \zb}{- 2 \pi i}$.  The world-sheet correlator admits a perturbation series in the curvature expansion,
\begin{equation}
    G^{tot}(z,\zb;S,T) =\sum_{k=0}^{\infty} \lambda^{-\frac{k}{2}} G^{tot,(k)}(z,\zb;S,T) \, ,
\end{equation}
which is bootstrapped up to second order curvature corrections \cite{Alday:2023mvu}\footnote{We have used the AdS/CFT dictionary, $R^4/\ell_s^4=\lambda$, where $R$ is the AdS curvature and $\ell_s$ is the string length, to express the large AdS curvature expansion in terms of the large-$\lambda$ expansion.}. For a general integrand $G^{tot,(k)}$, it is convenient to write down the following decomposition
\begin{align}  \label{eq:Gtot}
    G^{tot,(k)} = & \frac{1}{U^2} G^{(k)}(z ; S,T) + \frac{|z|^2}{S^2}  G^{(k)}\left(\frac{1}{z}; U,T \right) \nonumber \\
    &+ \frac{|1-z|^2}{T^2}G^{(k)}\left(\frac{z}{z-1};S,U\right) \, ,
\end{align}
where $U=-S-T$. 
This decomposition renders the crossing symmetry of $A(S,T)$ manifest. 

Altogether, \cref{EECfromAdSVS} provides a world-sheet formulation for the EEC at strong coupling. Taking this expression as our starting point, we will demonstrate that by interchanging the order of integrals, one can systematically derive the expression for the general coefficients $c_{k,n}$.

\vspace{0.8em}
\noindent{\bf Dealing with integrals.} While \cref{EECfromAdSVS} involves multiple layers of integration, each component admits a clear physical interpretation. We first recast the expression into the following form
\begin{equation}\label{EECfromI}
    \text{EEC}(\xi)=  \frac{1}{\lambda}\int_{-i\infty}^{+i\infty} \hspace{-0.4em}\frac{\mathrm{d}t}{2 \pi i} \int_{0}^{\infty}  \hspace{-0.4em} \mathrm{d \beta} \,{e^{-\beta}}  \beta^4 I(\beta\hat{t}) \, K(t,\xi) \, .
\end{equation}
Here, we have introduced the function $I(T)$, defined as
\begin{equation} \label{eq:relation}
\hspace{-0.56em}     I(T) =\!\! \int_{-i \infty}^{+i\infty} \hspace{-0.4em} \frac{\mathrm{d}s}{2 \pi i} \frac{\beta}{2 \sqrt{\lambda}} A(\beta \hat{s},\beta\hat{t}) = \!\!  \int_{-i \infty}^{+i\infty} \hspace{-0.4em}\frac{\mathrm{d}S}{2 \pi i} A(S,T) \, ,
\end{equation}
where we have identified $S=\beta \hat{s}$ and $T=\beta \hat{t}$, absorbing the scaling factors appropriately. This reformulation disentangles the specific contributions of each integral. {Since $I(T)$ is the central ingredient in computing the EEC, the relation \cref{eq:relation} establishes a direct connection between superstring amplitudes in AdS and the EEC in the dual CFT.} For stringy corrections, $I(T)$ can be expanded as a series in $\beta^i t^j$ in the large $\lambda$ expansion. The $\beta$-integral in \cref{EECfromI} then effectively acts as a rescaling, mapping $\beta^i \to \Gamma(i+5)$ and $t^j$ to the functions $Q_j(\xi)$. This mapping renders the relationship between the EEC and the strong coupling expansion of $I(T)$ transparent. The amplitude $A(S,T)$ admits a curvature expansion analogous to that of the world-sheet correlator; accordingly, we define $I^{(k)}(T)$ as the integral of the $k$-th order amplitude $A^{(k)}(S,T)$ over $S$. Finally, the determination of the coefficients $c_{k,n}$ is reduced to the small-$T$ expansion of $I(T)$.

It is worth commenting on the phenomenon of ``enhancement" characteristic of the EEC at strong coupling. Previous studies on the leading coefficient $c_{0,0}$ revealed that the result is enhanced to $4 \pi^2/\lambda$, exceeding the naive expectation from leading stringy corrections. Similar enhancements have been observed in higher-order large-$N$ corrections \cite{Chen:2024iuv}. We identify that this enhancement arises from the $S$-integration within $I(T)$. The Jacobian factor $\sqrt{\lambda}$, emerging from the transformation between Mellin variables and flat-space parameters, increases the stringy correction from order $\lambda^{-3/2}$ to $\lambda^{-1}$. This mechanism is universal across the entire stringy correction sector.

We now turn to the evaluation of the world-sheet and $S$-integrals. A significant simplification is achieved by interchanging the order of integration. The generic integral of interest takes the following form: 
\begin{align}\label{eq:Jjint}
    \mathcal{J}_{j}(T) = & \int \frac{\mathrm{d}S}{2\pi i} \int \mathrm{d}^2z  \, S^j |z|^{-2S-2} J(z,\zb;T)  \, ,
\end{align}
where $J(z,\bar{z};T)$ collects the remaining parts of the integrand of the AdS Virasoro-Shapiro amplitude. Note that terms involving a $U$ pole can be mapped to the same form via the shift $S\to S-T$. We evaluate the integral depending on whether the integer $j$ is negative or non-negative. 

\emph{Case 1: $j < 0$.} We can employ the Schwinger parametrization
\begin{equation*}
    S^j =\frac{1}{\Gamma(-j)}\int_{0}^{\infty} \mathrm{d}\alpha \, \alpha^{-j-1} \, e^{-\alpha S} \, .
\end{equation*}
Substituting this into $\mathcal{J}_j(T)$ and performing the $S$-integral yields a delta function constraint
\begin{equation*}
    \int_{-i \infty}^{+i\infty} \frac{\mathrm{d}S}{2\pi i} e^{-S (\ln|z|^2+\alpha)}  = \delta(\alpha+\ln|z|^2) \, .
\end{equation*}
Since the integration range for the Schwinger parameter is $\alpha \in [0, \infty)$, the support of delta function imposes the constraint $\ln|z|^2 < 0$, thereby restricting the spatial integral to the unit disk $|z|<1$ \footnote{We have independently verified this conclusion using the residue theorem. The convergence condition for the contour integral depends on the sign of $\ln|z|^2$. Assuming $\text{Re}(S)>0$ requires closing the contour in the left half-plane, which selects the region $|z|<1$. Conversely, assuming $\text{Re}(S)<0$ dictates closing the contour to the right, corresponding to $|z|>1$. While these domains are complementary, the sign reversal from the change in contour orientation cancels the sign difference between the radial integrals using the formal integral relation $\int_{0}^{1} dr \, r^{k} = - \int_{1}^{\infty} dr \, r^{k}$. Consequently, the final result is independent of the assumption regarding $\text{Re}(S)$.}. This allows us to derive explicit results for specific cases, such as $j=-1, -2$,
\begin{align}
\mathcal{J}_{-1}(T) = & \int_{|z|<1} d^2z \, |z|^{-2}  J(z,\zb;T) \, , \\
    \mathcal{J}_{-2}(T) = & \int_{|z|<1} d^2z \, |z|^{-2}  (-\ln|z|^2) J(z,\zb;T) \, . \label{eq:Jj-2}   
\end{align}

\emph{Case 2: $j \geq 0$.} Let us first rewrite the integrand using $|z|^{-2S-2} = |z|^{-2} e^{-S \ln|z|^2}$ and utilize the integral representation of delta function
\begin{equation*}
    \int_{-i \infty}^{+i\infty} \frac{\mathrm{d}S}{2\pi i} S^j e^{-S X} = (-1)^j \delta^{(j)}(X) \,.
\end{equation*}
Setting $X = \ln|z|^2$, the integral is localized on the unit circle $|z|=1$. To evaluate this, we switch to polar coordinates $z = r e^{i\theta}$. Using the property $\int \delta^{(j)}(X) f(X) dX = (-1)^j f^{(j)}(0)$, the integral reduces to a boundary integral over the unit circle
\begin{align}
    \mathcal{J}_{j}(T) = \oint_{|z|=1} {\mathrm{d} \theta \over 2\pi}   \calD^j J(z,\zb;T)   \, , \,\,\, {\rm with} \,\,\,\,\, \calD = {r \over 2}   \partial_r \, . 
\end{align}

We stress that the integral representations derived above remain valid for the general world-sheet correlation function $G^{(k)}(z,\bar{z};S,T)$. This establishes a systematic strategy for computing energy-energy correlations at higher orders of curvature corrections at strong coupling. Furthermore, when the contributions are combined, we observe that the calculation undergoes significant simplification. We will apply this approach to the flat-space contribution and the first curvature correction as concrete examples.

\vspace{0.8em}
\noindent{\bf Flat space contribution.} We begin by examining the flat space contributions $I^{(0)}(T)$, which governs the coefficients $c_{k,0}$. In this case, $G^{tot,(0)}=|z|^2/S^2$ in \cref{eq:Gtot} corresponds to $J(z,\zb;T)= |z|^2 |1-z|^{-2T-2}$ with $j=-2$ in \cref{eq:Jjint} \footnote{Note we should remove the supergravity amplitude $1/(S T U)$ for focusing the stringy effects.}. Identifying this with \cref{eq:Jj-2}, we evaluate the integral $I^{(0)}(T)$ as follows 
\begin{align} \label{eq:I0T}
   I^{(0)}(T) &= \int_{0}^1  \, \frac{r \, \mathrm{d}r  }{\pi}  (-2\ln r)\int_0^{2\pi} \, \mathrm{d} \theta \, |1-z|^{-2T-2}  \,  \nonumber \\ 
   &=\frac{1}{T^2} \left(\frac{\Gamma (1-2 T)}{\Gamma (1-T)^2} -1 \right) \,  .
\end{align}
The Taylor expansion of $I^{(0)}(T)$ for small $T$ reads
\begin{equation}
    I^{(0)}(T) = \zeta_2+ 2 \zeta_3 T + \frac{19}{4}\zeta_4 \, T^2  + \cdots \, .
\end{equation}
We observe that these coefficients align precisely with the leading contributions from the highest-spin sector \cite{Dempsey:2025yiv}. Indeed, these coefficients are uniquely fixed by the flat-space limit. Crucially, due to the relation $T=\beta\hat{t}$, the flat-space terms affect the sub-leading coefficients $c_{k,n}$ via a shift $t-4/3$, which should not be neglected.

Alternatively, exploiting the symmetry allows us to consider an equivalent integrand $J(z,\bar{z};T)=|1-z|^{-2T}/T^2$, with $j=0$ in \cref{eq:Jjint}. This corresponds to {\it case 2}, and the resulting integral is localized on the boundary of a disk  
\begin{equation} \label{eq:I0-newform}
    I^{(0)}(T) = \frac{1}{T^2}  \int_0^{2\pi} \frac{\mathrm{d} \theta }{2\pi}  \left[ 4 \sin^2\left({\theta \over 2 }\right)
    \right]^{-T}  \, .
\end{equation}
Note that the $1/T^2$ term in small-$T$ expansion should be removed, since it corresponds to the supergravity contribution, which we have excluded by focusing on stringy effects. 
It is worth noting that this alternative form can also be obtained by interpreting the energy detector  as a shockwave perturbation of the dual AdS geometry \cite{Hofman:2008ar}, and the two-point correlator of the corresponding shockwave vertex operators then reproduces the expression in \cref{eq:I0-newform} (see also \cite{Dempsey:2025yiv}).

\vspace{0.8em}
\noindent{\bf First curvature correction.} We now turn to the first curvature correction. This contribution can be decomposed into bulk and boundary integrals. For completeness, the full expression for the first curvature correction to the AdS Virasoro-Shapiro amplitude~eq.(A2), is provided in the Appendix (A) \footnote{See Supplemental Material for details on first curvature correction to the EEC, which includes Refs. \cite{Goncharov:1998kja,Goncharov:2001iea,Brown:2004ugm,Duhr:2019tlz,Alday:2023jdk,Alday:2023mvu}.}, where details of the calculation are presented. We observe that restricting the integration domain precipitates dramatic cancelations among complicated polylogarithmic structures; the resulting concise integral representation hints at the intrinsic simplicity of the EEC. In summary, the final integrated result is organized into four distinct functions
\begin{equation}
    I^{(1)}(T) = \sum_{i=1}^4 I^{(1)}_i(T) \, ,
\end{equation}
where $I^{(1)}_1(T)$ and $I^{(1)}_2(T)$ originate from the bulk integration, while $I^{(1)}_3(T)$ and $I^{(1)}_4(T)$ arise from the boundary contributions, and their explicit world-sheet integral representations are given in the Appendix (A). We summarize the results as follows.  

The bulk contributions, supported on the disk $|z|<1$, are given by
\begin{align}
    I^{(1)}_1(T)+ I^{(1)}_2(T) = - \frac{T}{9} \frac{\partial ^3}{\partial T^3} \frac{2 \Gamma (-2 T)}{\Gamma (1-T)^2}  \, .
\end{align}
The small-$T$ expansion yields a well-behaved series where the coefficients are in terms of zeta values. The remaining contributions are localized on the boundary $|z|=1$ and involve polylogarithms arising from the world-sheet correlator expansion
\begin{align}
    I^{(1)}_3 &=2 \, \zeta_3 \frac{ \Gamma (-2 T-1)}{\Gamma (-T)^2} + F(T) \, ,\\
    I^{(1)}_4 &= -\frac{\partial ^3}{\partial T^3} \left(\frac{5 \Gamma (1-2 T)}{18 \Gamma (1-T)^2}\right)-F(T-1) \, ,
\end{align}
where $F(T)$ is defined by an infinite sum
\begin{align}
    F(T)&=\sum_{n=1}^{\infty}\frac{2 \sin (\pi  T) \Gamma (-2 T-1) \Gamma (n+T+1)}{\pi  n^3 \Gamma (n-T)} \, . \label{eq:FT}
\end{align}
Although $F(T)$ can be resumed into a generalized hypergeometric function, the infinite sum representation is more advantageous for the Taylor expansion in $T$.

Collecting all contributions, we find that the complete first curvature correction is captured by the following compact form 
\begin{align} \label{eq:I1T}
    I^{(1)}(T)=&  \, \frac{1}{3 T^4} \frac{\partial }{\partial T}T^5\frac{\partial ^2}{\partial T^2} \frac{\Gamma (-2 T)}{ \Gamma (1-T)^2}\nonumber \\
    &+\frac{ 2 \zeta_3\,  \Gamma (-2 T-1)}{\Gamma (-T)^2}  + F(T) -F(T-1)  \, .
\end{align}
We stress again that $I^{(1)}(T)$ appears to be much simpler than the corresponding AdS Virasoro-Shapiro amplitude.

Evaluating \cref{eq:EECxi} requires the small-$T$ expansion of $I^{(1)}(T)$, see Appendix (B) \footnote{See Supplemental Material for series expansion of $I(T)$.} for technical details. Remarkably, all contributions involving the Euler-Mascheroni constant cancel out, revealing a structure governed by multiple zeta values. The first few terms are given by
\begin{align}
     I^{(1)}(T)=\zeta_2-\frac{10}{3} \zeta_3 +   \left(\! -2 \zeta_2 +4 \zeta_3-\frac{53}{2} \zeta_4 \right) T + \cdots \, , \nonumber
\end{align}
where we have dropped  the singular term in the small-$T$ expansion, which is associated with the supergravity contribution that we should exclude. Finally, combining the contributions from the flat-space results and first curvature corrections, we can determine all the coefficients $c_{k,0}$ and $c_{k,1}$ in \cref{eq:EECxi}. In particular, $c_{k,0}$ can be straightforwardly read off from \cref{eq:I0T}, yielding 
\begin{align} \label{eq:ck0}
    c_{0,0}= 24 \zeta_2 \, , \quad c_{1,0}= 120 \zeta_3 \, , \quad c_{2,0}= 855 \zeta_4 \, ,~~ \cdots \, , 
\end{align}
and for $c_{k,1}$, we derive the following results, 
\begin{align} \label{eq:c11}
    c_{1,1}&= 24 \zeta_2 - 240 \zeta_3 \, ,\\
    c_{2,1}&=-120 \zeta_2 +240 \zeta_3 -3870 \zeta_4 \, , ~~ \cdots \, ,    \label{eq:c21} 
\end{align}
and similar results for higher order terms. 
This gives the strong coupling expansion of the EEC, 
\begin{align}
    \text{EEC}&(\xi)
    =1+ \frac{24\zeta_2}{\lambda} Q_{0}(\xi) \\ 
    &+ \frac{(24 \zeta_2 - 240 \zeta_3 )\, Q_{0}(\xi)+ 120 \zeta_3\, Q_{1}(\xi)}{\lambda^{3/2}} + \cdots\, . \nonumber
\end{align}
As a consistency check, we performed an independent calculation for $c_{0,0}$, $c_{1,0}$ and $c_{1,1}$ using the low-energy expansion method in Appendix (C) \footnote{See Supplemental Material for low-energy expansion, which includes Refs. \cite{Alday:2022xwz}.}, which yields exactly identical results~\footnote{We note our result for $c_{1,1}$ disagrees with that in the ref.~\cite{Dempsey:2025yiv}, corresponding to the $\lambda^{-3/2}$ term in eq.~(4.6) of that reference. Converting to our basis, their expression implies $c_{1,1}= 48 \zeta_2 - 240 \zeta_3$. We find that this discrepancy arises from performing the $s$-integration prior to the $\delta$-summation in \cite{Dempsey:2025yiv}, which omits a certain subtle contribution (more precisely, a term with a $0/0$ structure). We have explicitly verified that including such contribution restores precise agreement. We thank Alexander Zhiboedov for very helpful discussions on this point.}. 

\vspace{0.8em}

\noindent{\bf Discussion.} In this Letter, we have established a rigorous connection between the EEC in strongly coupled $\mathcal{N}=4$ SYM and the world-sheet description of the AdS Virasoro-Shapiro amplitude. By reformulating the problem from Mellin space into the integration over the string moduli space, we demonstrated that the computation of stringy corrections to the EEC reduces to evaluating integrals over a unit disk and its boundary on the Riemann sphere. In practice, this powerful construction leads to an extremely efficient way of computing the EEC in $\mathcal{N}=4$ SYM at strong coupling. We demonstrated this by explicitly computing the flat space and first curvature correction to the EEC and obtained finite results. 

The results of this Letter open up a variety of interesting directions for future work. A natural extension is to consider the second order of the higher-curvature correction \cite{Alday:2023mvu}. Such higher-order results would also allow for a more precise comparison with the bootstrap bounds \cite{Dempsey:2025yiv}.  Another direction is to study the EECs in $\mathcal{N}=4$ SYM involving operators with higher dimensions \cite{Chicherin:2023gxt} for the relevant correlators \cite{Fardelli:2023fyq}. 
Perhaps an even more ambitious goal is to bootstrap the EEC directly. The observed structural simplicity of the EEC suggests it may be feasible to determine these corrections without referencing the AdS Virasoro-Shapiro amplitude. 

Furthermore, $\mathcal{N}=4$ SYM is known to enjoy S-duality \cite{Montonen:1977sn, Witten:1978mh, Osborn:1979tq}. It has been shown that the first few orders of stringy corrections to the correlator can be determined exactly in terms of known non-holomorphic modular functions, which incorporate all perturbative and instanton contributions and manifest S-duality \cite{Chester:2019jas, Chester:2020vyz}. It would be interesting to study the modular properties of the EEC and its non-perturbative instanton effects. 

Finally, our world-sheet formulation may offer new insights into the back-to-back limit ($\xi\to1$) of EEC. Its singular behavior is governed by Sudakov logarithms \cite{Collins:1981uk,Moult:2018jzp,Korchemsky:2019nzm,Chen:2023wah}, which are effectively described by the geometry of Wilson lines \cite{Korchemsky:1987wg,Chen:2025ffl}. Intriguingly, a parallel geometric description exists at strong coupling, where correlation functions in the light-like limit are known to map directly to null polygonal Wilson loops \cite{Alday:2010zy}. The prominence of Wilson loops at both weak and strong coupling suggests a universal physical picture: the infrared dynamics of energy flow may be effectively described by specific Wilson loop configurations, even in the intermediate stringy regime. It is tantalizing to speculate that our world-sheet results could be reorganized into such an effective ``stringy Wilson loop" description. Identifying these dominant operators that control this regime would be a crucial step toward constructing an effective string theory for jet physics.

\vspace{0.5cm}

	\begin{acknowledgments}
	\noindent{\bf Acknowledgments.}
	   The authors would like to thank Hao Chen, Francesco Galvagno, Alessandro Georgoudis, Max Jackson, Murat Kologlu, Rodolfo Russo, Joao Vilas Boas, Alexander Zhiboedov, and Huaxing Zhu for valuable discussions on related topics. 
       LR is supported by the Royal Society via a Newton International Fellowship. 
       BW is supported by the National Natural Science Foundation of China under Grant No.~124B2095 and Grant No.~12175197. CW is supported by a Royal Society University Research Fellowship,  URF$\backslash$R$\backslash$221015 and a STFC Consolidated Grant, ST$\backslash$T000686$\backslash$1 ``Amplitudes, strings \& duality".
	\end{acknowledgments}
	
	\bibliography{refs}

\widetext
\begin{center}
	\textbf{\large Supplemental Materials}
\end{center}
\appendix
\section{Details on first curvature correction to the EEC}\label{app:svmpl}

The first curvature correction to the AdS Virasoro-Shapiro amplitude can be expressed as a moduli space integral over \textit{multiple polylogarithms}(MPLs), denoted by $L_{w}(z)$. These functions are labeled by a ``word'' $w$ formed by a set of complex variables $(a_1,a_2,\ldots,a_n)$ known as ``letters''. They defined recursively through iterated integrals \cite{Goncharov:1998kja,Goncharov:2001iea}
\begin{align}
    L_{a_1a_2\ldots a_n}(z)=\int_0^z\frac{\mathrm{d}t}{t-a_1}L_{a_2a_3\ldots a_n}(t)\,, 
\end{align}
where the length of the word $w$ is called the transcendental weight, with $L_{\varnothing}(z):=1$.
In our case, we consider only words composed of letters  $a \in \{0,1\}$. 

The MPLs are, by definition, multi-valued functions that depend on the choice of contour at each integral. Therefore, we shall use single-valued multiple polylogarithms (SVMPLs) \cite{Brown:2004ugm} as the moduli space integrand for string amplitudes. This involves a map $\text{sv:}\;L \to {\cal L}$ such that ${\cal L}_{w}$ becomes a single-valued linear combination of $L_{w'}(z)L_{w''}(\bar z)$ with $w'+w'' = w$ preserving the weight. The map can be conveniently implemented using the {\tt Mathematica} package PolyLogTools \cite{Duhr:2019tlz}. In the package, SVMPL is denoted by \texttt{cG}, and we define $\mathcal{L}_{a_1 a_2 \cdots a_n}(z)=\mathtt{cG[a_1,a_2,\cdots,a_n,z]}$. 

For the first curvature corrections, the SVMPLs of our interest have transcendental weight 3~\cite{Alday:2023jdk,Alday:2023mvu}.
For convenience, we provide the explicit representations of these SVMPLs in terms of MPLs below,
\begin{align*}
    \mathcal{L}_{000}(z)&=L_{0}(z)L_{00}(\zb) +L_{0}(\zb)L_{00}(z) +L_{000}(z)+L_{000}(\zb) \, , \\
    \mathcal{L}_{111}(z)&=L_{1}(z)L_{11}(\zb) +L_{1}(\zb)L_{11}(z) +L_{111}(z)+L_{111}(\zb) \, , \\
    \mathcal{L}_{100}(z)&=L_{1}(z)L_{00}(\zb) +L_{0}(\zb)L_{10}(z) +L_{100}(z)+L_{001}(\zb) \, , \\
    \mathcal{L}_{010}(z)&=L_{0}(z)L_{01}(\zb) +L_{0}(\zb)L_{01}(z) +L_{010}(z)+L_{010}(\zb) \, , \\
    \mathcal{L}_{001}(z)&=L_{0}(z)L_{10}(\zb) +L_{1}(\zb)L_{00}(z) +L_{001}(z)+L_{100}(\zb) \, , \\
    \mathcal{L}_{011}(z)&=L_{0}(z)L_{11}(\zb) +L_{1}(\zb)L_{01}(z) +L_{011}(z)+L_{110}(\zb) \, , \\
    \mathcal{L}_{101}(z)&=L_{1}(z)L_{10}(\zb) +L_{1}(\zb)L_{10}(z) +L_{101}(z)+L_{101}(\zb) \, , \\
    \mathcal{L}_{110}(z)&=L_{1}(z)L_{01}(\zb) +L_{0}(\zb)L_{11}(z) +L_{110}(z)+L_{011}(\zb) \, .
\end{align*}
The integrand of the first curvature correction of the AdS Virasoro-Shapiro amplitude then reads \cite{Alday:2023jdk,Alday:2023mvu},
\begin{equation}\label{eq:AdSVS1}
    G^{tot,(1)} =  \frac{2}{3}\frac{T}{S+T} G^{ST}_1 + \frac{1}{6}G^{ST}_2 +|1-z|^2 \left( \frac{2S}{3T} G^{SU}_1 + \frac{1}{6}G^{SU}_2\right) +|z|^2 \left( \frac{2T}{3S} G^{UT}_{1} + \frac{1}{6} G^{UT}_{2} \right) \, ,
\end{equation}
where the coefficient functions $G$ are given by the following combinations of the SVMPLs
\begin{align*}
    G^{ST}_1&= \mathcal{L}_{000}(z)-\mathcal{L}_{001}(z)-\mathcal{L}_{010}(z)+\mathcal{L}_{011}(z)-\mathcal{L}_{100}(z)+\mathcal{L}_{101}(z)+\mathcal{L}_{110}(z)-\mathcal{L}_{111}(z) \, , \\
    G^{ST}_2&= -4 \mathcal{L}_{000}(z)+2 \mathcal{L}_{001}(z)-\mathcal{L}_{010}(z)-2 \mathcal{L}_{011}(z) + 2 \mathcal{L}_{100}-5 \mathcal{L}_{101}(z)-2 \mathcal{L}_{110}(z) \, , \\
    G^{SU}_1&=  \mathcal{L}_{000}(z) \, ,  \quad 
    G^{SU}_2= 2 \mathcal{L}_{001}(z)+5 \mathcal{L}_{010}(z)-5 \mathcal{L}_{011}(z) + 2 \mathcal{L}_{100} -5 \mathcal{L}_{101}(z) -5 \mathcal{L}_{110}+10 \mathcal{L}_{111}(z) \\
    G^{UT}_1 &= \mathcal{L}_{111}(z) \, , \quad  
    G^{UT}_2 = 10 \mathcal{L}_{000}(z)  -5 \mathcal{L}_{001}(z)-5\mathcal{L}_{010}(z)+2 \mathcal{L}_{011}(z)-5 \mathcal{L}_{100}+5 \mathcal{L}_{101}(z)+2 \mathcal{L}_{110}+12 \zeta_3 \, .
\end{align*}
Based on the decomposition into bulk and boundary contributions discussed in the main text, we evaluate the associated integrals explicitly. The first bulk integral, supported on the unit disk $|z|<1$ from $G^{ST}_1$, is evaluated as
\begin{align}  
    I^{(1)}_1(T)=  \frac{T}{9} \int_{|z|<1}  d^2z \, |z|^{2T-2} \,  |1-z|^{-2T-2} \, \log^3 \frac{|z|^2}{|1-z|^2}  = -\frac{T}{9}\frac{\partial ^3}{\partial T^3}  \frac{\Gamma (-2 T)}{\Gamma (1-T)^2} \, . \nonumber
\end{align}
The second bulk integral, supported on the unit disk $|z|<1$ from $G^{UT}$, is evaluated as
\begin{align}  
    I^{(1)}_2(T) &=  \frac{T}{9} \int_{|z|<1}  d^2z \, |1-z|^{-2T-2} \,  \log^3|1-z|^2  = -\frac{T}{9}\frac{\partial ^3}{\partial T^3}  \frac{\Gamma (-2 T)}{\Gamma (1-T)^2} \, .\nonumber
\end{align}
Note that these two contributions yield the same result. The remaining contributions localize on the boundary $|z|=1$. These reduce to one-dimensional contour integrals involving polylogarithms, which can be evaluated in terms of infinite sums of Gamma functions. The contribution from $\frac{1}{6}(G^{ST}_2+G^{UT}_2)$ reads
\begin{align}  
    I^{(1)}_3(T) &=  \oint_{|z|=1}  \frac{\mathrm{d}\theta}{2\pi}  |1-z|^{-2T-2} \, \left(2 \zeta_3 -\text{Li}_3(z)-\text{Li}_3(\zb) \right) = \frac{2 \zeta_3\, \Gamma (-2 T{-}1)}{\Gamma (-T)^2} + \sum_{n=1}^{\infty}\frac{2 \sin (\pi  T) \Gamma (-2 T{-}1) \Gamma (n{+}T{+}1)}{\pi  n^3 \Gamma (n{-}T)} \, . \nonumber
\end{align}
Notice that $G^{SU}_1$ does not contribute to the boundary integral. The final integral for $\frac{1}{6}G^{SU}_2$ is
\begin{align}  
    I^{(1)}_4(T) &= \! \oint_{|z|=1} \frac{\mathrm{d}\theta}{2\pi} \,  |1{-}z|^{-2T} \left( \text{Li}_3(z)+\text{Li}_3(\zb)+\frac{5}{18} \log^3 |1{-}z|^2 \right) =  \! \sum_{n=1}^{\infty}\frac{2 \sin (\pi  T) \Gamma (1{-}2 T) \Gamma (n {+} T)}{\pi  n^3 \Gamma (n{-}T{+}1)}  - \frac{\partial ^3}{\partial T^3} \frac{5 \, \Gamma (1 {-} 2 T)}{18\, \Gamma (1 {-} T)^2} \, . \nonumber
\end{align}
We note that the summation appearing in $I^{(1)}_3(T)$ involves the function $F(T)$ defined in the main text, while the sum in $I^{(1)}_4(T)$ involves the shifted function $F(T-1)$. Putting all the contributions together, we have
\begin{align} \label{aeq:I1}
    I^{(1)}(T)=  \frac{ 2 \zeta_3\, \Gamma (-2 T{-}1)}{\Gamma (-T)^2} +\frac{1}{3 T^4} \frac{\partial }{\partial T}T^5\frac{\partial ^2}{\partial T^2} \frac{\Gamma (-2 T)}{ \Gamma (1{-}T)^2}+ \frac{\sin (\pi T)\, \Gamma (-2 T{-}1) }{\pi} \sum_{n=1}^{\infty}\frac{2 \left(n^2{+}3T^2{+}2T\right)   \Gamma (n{+}T)}{  n^3 \Gamma (n{-}T{+}1)} \, .
\end{align}

\section{Series expansion of $I(T)$}
\label{app:Taylor-exp}

In this appendix, we detail the algorithm for the series expansion of $I^{(1)}(T)$. 
This can be achieved by recasting ratios of Gamma functions into the following exponential forms,
\begin{align}
   \frac{\Gamma (1-2 T)}{ \Gamma (1-T)^2} = \exp \left(\sum _{m=2}^{\infty} \frac{\left(2^m{-}2\right) T^m \zeta_m}{m}\right) \, , 
   \end{align}
   and for the function $F(T)$ we use
   \begin{align}
    \frac{\Gamma (-2 T{-}1) \Gamma (n{+}T{+}1)}{\Gamma (n-T)} = \frac{n}{2T}\exp \left(\sum _{m=1}^{\infty}  \frac{d_m T^m}{m} \right) \, , 
\end{align}
where the coefficients $d_m(n)$ depend on generalized harmonic numbers $H_m(n-1)=\sum_{k=1}^{n-1} k^{-m}$,
\begin{align}  \label{eq:dm}
d_m(n)=\left(1-(-1)^m\right) H_m(n-1)+(-1)^m \left(2^m-\frac{1}{n^m}\right) +\left(2^m+(-1)^m-1\right) \zeta_m \, . 
\end{align}
The central challenge lies in evaluating infinite sums over products of harmonic numbers in $F(T)$, which requires the properties of nested harmonic numbers and their connection to multiple zeta values (MZVs). To evaluate these systematically, we utilize the algebra of \textit{nested harmonic sums}. Let us define
\begin{equation} 
H_{k_1, k_2, \dots, k_r}(N) = \sum_{n=1}^{N} \frac{1}{n^{k_1}} H_{ k_2, \dots, k_r}(n-1) \, ,
\end{equation}
with $ H_{\varnothing}(N)=1$. Here, the sequence $k_1, k_2 ,\cdots , k_r$ constitutes the index vector, where $r$ denotes the \textit{depth} and $w=\sum_{i=1}^{r}k_i$ denotes the \textit{weight} (transcendental weight).

The product of two nested harmonic sums obeys the quasi-shuffle (stuffle) algebra. Let $\vec{a} = (a, \vec{u})$ and $\vec{b} = (b, \vec{v})$ be two index sequences, where $a$ and $b$ denote the first indices and $\vec{u}, \vec{v}$ denote the remaining vectors. The product relation is defined recursively by
\begin{equation}
    H_{\vec{a}}(N) H_{\vec{b}}(N) = H_{\vec{a} * \vec{b}}(N) \, ,
\end{equation}
where the quasi-shuffle product $*$ is given by
\begin{equation}
    (a, \vec{u}) * (b, \vec{v}) = \left(a, \vec{u} * (b, \vec{v})\right) + \left(b, (a, \vec{u}) * \vec{v}\right) + \left(a+b, \vec{u} * \vec{v}\right) \, .
\end{equation}
The final term, characteristic of the quasi-shuffle algebra, accounts for the contribution where the summation indices coincide. For example, at depth 1, this yields $H_a(N) H_b(N) = H_{a,b} (N)+ H_{b,a} (N) + H_{a+b} (N)$.

By iteratively  applying these shuffle relations, any product of harmonic numbers appearing in the small-$T$ expansion of $F(T)$ can be decomposed into a linear combination of nested harmonic sums. In the limit $N \to \infty$, these sums converge to multiple zeta values (MZVs), defined by
 \begin{equation}
    \zeta_{k_1, \dots, k_r} = \lim_{N \to \infty} H_{k_1, \dots, k_r}(N) = \sum_{1 \le n_r < \dots < n_1 < \infty} \frac{1}{n_1^{k_1} \dots n_r^{k_r}} \, ,
\end{equation}
provided $k_1 > 1$ to ensure convergence. As an illustrative example, we have 
\begin{align}
    \sum_{n=1}^{\infty} \frac{H_1(n-1) H_3(n-1)}{n^2} =\sum_{n=1}^{\infty}  \frac{H_{1,3}(n-1)+H_{3,1}(n-1)+H_{4}(n-1)}{n^2}= \zeta_{2,1,3} + \zeta_{2,3,1} + \zeta_{2,4} =\frac{167}{48} \zeta_6-\frac{3 }{2} \zeta_3^2 \, .
\end{align}
These expressions ensure that the summation over $n$ yields sums over nested harmonic numbers via shuffle algebra, which evaluates to multiple zeta values. Importantly, this guaranties the complete cancellation of the Euler-Mascheroni constant $\gamma_E$ in the final small-$T$ expansion, which provides a non-trivial consistency check of the result.  

\section{Energy-energy correlator from low-energy expansion} \label{app:low-energy}

In this appendix, we rederive some of the results of the EEC in main text independently using  low-energy expansion of the AdS Virasoro-Shapiro amplitude, which may serve as an consistency check. 
Recall the Mellin amplitude $M(\hs, \htt)$ can be expressed as 
\begin{align}
    M(\hs, \htt) = M^{\rm SUGRA}(\hs, \htt) + \frac{1}{\lambda^{3/2}} \sum_{\ell=0}^{\infty} \lambda^{-\ell/2} M^{(\ell)}(\hs, \htt)\, ,
\end{align}
$M^{(0)}(\hs, \htt)$ is the flat-space contribution, $M^{(1)}(\hs, \htt)$ is the first curvature correction, etc.. For computing the EEC, we expand it in small $\htt$, 
\begin{align}
    M^{(\ell)}(\hs, \htt) =\sum_{j=0}^{\infty} M^{(\ell)}_j (\hs)\, \htt^j \,, 
\end{align}
and use the definition, we have
\begin{align}
    \text{EEC}(\xi)
    &=1+ \sum_{\ell=0}^{\infty} \sum_{j=0}^{\infty} \lambda^{-\frac{\ell}{2}-1} \, \hat{c}_{j,\, \ell} \, \widehat{Q}_{j}(\xi) \, , 
\end{align}
where the first term (i.e. $1$) arising from supergravity contribution $M^{\rm SUGRA}(\hs, \htt)$, and $\widehat{Q}_{n}(\xi)$ relevant for stringy effects are defined as
\begin{align}
    \widehat{Q}_{j}(\xi) = \int_{-i\infty}^{+i\infty} \frac{\mathrm{d}t}{2 \pi i}\, \htt^{\, j} \, K(t,\xi) \, ,
\end{align}
and the coefficients $\hat{c}_{j,\, \ell}$ is given by 
\begin{align}
    \hat{c}_{j,\, \ell} = \int_{-i \infty}^{+ i \infty} {\mathrm{d} \hs \over 2\pi i }\,M_j^{(\ell)}(\hs) \, . 
\end{align}
It is straightforward to see that $\widehat{Q}_{j}(\xi)$ is related to $Q_{n}(\xi)$ in a simple manner, from which one can also relate ${c}_{k,n}$ to $\hat{c}_{j,\ell}$. For example, 
\begin{align} \label{eq:relations}
 {c}_{k,0} = 2^{-k}\, \hat{c}_{k, 0}  \, ,  \qquad  c_{k+1, 1} = 2^{-k}\, \left( \hat{c}_{k, 1} -\frac{2(k+1)}{3} \hat{c}_{k+1, 0}\right)\, . 
\end{align}

We now compute $\hat{c}_{j,\, \ell}$ from the low-energy expansion of the Mellin amplitude. From the flat-space Virasoro-Shapiro amplitude,  we have, for example, 
\begin{align}
   M^{(0)}_{0}(\hs) &= 2 \sum_{n=0}^{\infty} \Gamma(2n+6) \, \zeta_{2n+3} \, \hs^{2n} \, ,  \\  
   M^{(0)}_{1}(\hs) &= -2 \sum_{n=0}^{\infty}  \Gamma(2 n+7) \sum_{i=1}^{n} \zeta_{2 i + 1} \zeta_{2n+3-2 i} \, \hs^{2n} + 2 \sum_{n=1}^{\infty}  n \, \Gamma (2 n+6) \, \zeta_{2n+3} \,\hs^{2 n-1} \, . 
\end{align}
These asymptotic series can be made sense through Borel resummation. We will apply a modified Borel transform \cite{Arutyunov:2016etw} (see \cite{Dorigoni:2021guq} for its application for corerelators in $\mathcal{N}=4$ SYM), using the identity
\begin{align} \label{eq:zeta-id}
  \Gamma(n {+} 1)\,  \zeta_n= 2^{n-1}  \int_0^{\infty} dw \frac{w^n}{\sinh^2(w)}  \, . 
\end{align}
Apply the resummation to $ M^{(0)}_{0}(\hs)$, we then obtain 
\begin{align}
    \hat{c}_{0,0} = - \int_0^{\infty} \mathrm{d} w \int_{-i \infty}^{+ i \infty} {\mathrm{d} \hs \over 2\pi i}  \frac{32 w^3 \left(24 \hs^4 w^4-18 \hs^2
   w^2+5\right)}{\left(4 \hs^2 w^2-1\right)^3 \sinh^2(w)}= 4 \pi^2 \, . 
\end{align}
Similarly, we can express each $\zeta$-value in $M^{(0)}_{1}(\hs)$ using \cref{eq:zeta-id}, and after resummation we obtain a well-defined three-fold integral, from which we obtain 
\begin{align}
    \hat{c}_{1, 0} = 240\, \zeta_3  \, . 
    \end{align}
We may apply the same procedure to higher order terms arising from the flat-space amplitude as well as the curvature corrections, as those described in the above. For the later, using the results of \cite{Alday:2022xwz} (in particular eqs.(1.1) and (3.21) in that reference), we obtain the first curvature correction, 
\begin{align}
    \hat{c}_{0, 1} =  -80 \, \zeta_3 +24 \, \zeta_2 \, . 
    \end{align}
Using the relations \cref{eq:relations}, we find the results obtained from the low-energy expansion in a perfect agreement with results in the main text. 
\end{document}